\documentclass[aps,pra,twocolumn,superscriptaddress,floatfix,nofootinbib,showpacs,longbibliography]{revtex4-1}

\usepackage{xcolor}
\usepackage[utf8]{inputenc}  
\usepackage[T1]{fontenc}     
\usepackage[british]{babel}  
\usepackage[sc,osf]{mathpazo}
\usepackage[scaled=0.86]{berasans}  
\usepackage[colorlinks=true, citecolor=blue, urlcolor=blue]{hyperref}  
\usepackage[babel]{microtype}  
\usepackage{amsmath,amssymb,amsthm,bm,amsfonts,mathrsfs,bbm} 

\usepackage{xspace}  
\usepackage{pgf,tikz}
\usepackage{xcolor}
\usepackage{multirow}
\usepackage{array}
\usepackage{bigstrut}
\usepackage{braket}
\usepackage{color}
\usepackage{natbib}
\usepackage{multirow}
\usepackage{mathtools}
\usepackage{float}
\usepackage[caption = false]{subfig}
\usepackage{xcolor,colortbl}
\usepackage{color}

\newcommand{\be}{\begin{equation}}
\newcommand{\ee}{\end{equation}}
\newcommand{\ba}{\begin{eqnarray}}
\newcommand{\ea}{\end{eqnarray}}

\newtheorem{definition}{Definition}
\newtheorem{proposition}{Proposition}

\def\>{\rangle}
\def\<{\langle}

\begin{document}
\title{Operational Characterization of Multipartite Nonlocal Correlations}

\author{Sagnik Dutta}
\affiliation{Department of Physical Sciences, IISER Kolkata, Mohanpur 741246, West Bengal, India}

\author{Amit Mukherjee}
\affiliation{S.N. Bose National Center for Basic Sciences, Block JD, Sector III, Salt Lake, Kolkata 700098, India.}

\author{Manik Banik}
\affiliation{School of Physics, IISER Thiruvananthapuram, Vithura, Kerala  695551, India.}

\begin{abstract}
Nonlocality, one of the most puzzling features of multipartite quantum correlation, has been identified as a useful resource for device-independent quantum information processing. Motivated by the resource theory of quantum entanglement recently an operational framework have been proposed by Gallego {\it et al.}  [\href{https://doi.org/10.1103/PhysRevLett.109.070401}{Phys. Rev. Lett. {\bf 109}, 070401 (2012)}] and Bancal {\it et al.} [\href{https://doi.org/10.1103/PhysRevA.88.014102}{Phys. Rev. A {\bf 88}, 014102 (2013)}] that characterizes the nonlocal resource present in multipartite quantum correlations. While the bipartite no-signaling correlations allows a dichotomous classification -- local \textit{vs.} nonlocal, in multipartite scenario the authors have shown existence of several types of nonlocality that are inequivalent under the proposed operational framework. In this work we present a finer characterization of multipartite no-signaling correlations based on the same operational framework. We also clarify a statement in Gallego	{\it et al.}’s work that could be misinterpreted and make the conclusions of that work more precise here.
\end{abstract}

\maketitle
\section{Introduction}\label{intro}
Nonlocality captures one of the important characteristic aspects of multipartite quantum systems. John S. Bell, in his seminal work \cite{Bell:1964}, proved that a composite quantum system, prepared in suitable entangled state, can exhibit input-output correlations that can not be explained within the {\it local realistic} world view of classical physics \cite{Bell:1964,Bell:1966} (see also \cite{Brunner:2014,Mermin1993}). Bell considered quantum system with two spatially separated subsystems and termed the joint input-output correlation {\it locally-causal} if it is product of input-output probabilities for the individual parties or convex mixture of individual input-output probabilities. He derived an empirically testable inequality whose violation establishes {\it nonlocal} nature of the correlation. Later, Svetlichny initiated the study of nonlocal correlations for multipartite quantum systems that involved more than two subsystems \cite{Svet}. Following an apparently natural mathematical generalization of Bell he characterized multipartite correlations into three types - {\it fully local}, {\it bilocal}, and {\it genuine nonlocal}. He also came up with an empirical test to certify genuine nonlocality of a correlation. More recently, however, several groups of researchers identified that Svetlichny's definition is stronger than needed, and suggested that the definition be modified to optimally capture the 	notion of genuineness in perfect operational sense \cite{Gallego2012,Barrett2012}. They have also proposed a novel operational framework to address this issue.

The framework by Gallego \textit{et al.} \cite{Gallego2012} is well motivated from resource theoretic perspective where the set of relevant free operations is identified as \textit{wirings and classical communication prior to the inputs} (WCCPI). This framework introduces a hierarchical classification of multipartite correlations: set of no-signalling bilocal (NSBL) correlations $\subsetneq$ set of time-ordered bilocal correlations (TOBL) $\subsetneq$ set of general bilocal (BL) correlations $\subsetneq$ set of multipartite no-signalling (NS) correlations. The set BL is identical to the set of bilocal correlations as identified by Svetlichny and the correlations lying outside this set are called genuinely nonlocal. However, as pointed out in \cite{Gallego2012}, a correlation even within the set BL may exhibit unexpected behaviour under WCCPI protocol. This consequently indicates genuine nonlocal nature of those BL correlations and therefore indicates that the framework of Svetlichny be improved to	give a finer characterisation. In this work we revisit the framework of Gallego \textit{et al.} While exploring their work we find that their framework is operationally perfect, but also that	we should clarify some of their claims regarding the correlations presented there. Interestingly, we find that a more critical analysis of Ref. \cite{Gallego2012}, in fact, introduces new classes of correlations lying in between the sets TOBL and BL. Our work thus can be viewed as culmination of the novel operational framework by  Gallego \textit{et al.} for characterizing the nonlocal correlations in multipartite scenario. The article is organized as follows: in section (\ref{framework}) we recall the operational frameworks for multipartite nonlocal correlations that have already been developed in the literature, in section (\ref{result}) we present the main contribution of our work, and finally we put our conclusions in section (\ref{discussion}).    

\section{Framework}\label{framework}
Consider $n$ spatially separated parties. The input for $k^{th}$ party is denoted by $i_k$ and the corresponding outcome by $o_k$, with values taken from some finite set $\mathcal{I}_k$ and $\mathcal{O}_k$ respectively; $k\in\{1,\cdots,n\}$. An $n$-partite correlation is a joint input-output probability distribution $P:=\{p(o_1\cdots o_n|i_1\cdots i_n)~|~p(o_1\cdots o_n|i_1\cdots i_n)\ge0~\forall~i_k\in\mathcal{I}_k~\&~o_k\in\mathcal{O}_k;~\sum_{k}\sum_{o_k}p(o_1\cdots o_n|i_1\cdots i_n)=1~\forall~i_k\in\mathcal{I}_k\}$. Such a correlation is called no-signaling if any non-empty subgroup of the parties can not change the marginal outcome probabilities for the remaining parties by changing their choice of inputs. Set of all NS correlations we will denote as $\mathcal{P}^{NS}_n$ or simply as $\mathcal{P}^{NS}$ when number of parties are not relevant to mention. Study of quantum nonlocality identifies physically motivated different hierarchical subsets of correlations in $\mathcal{P}^{NS}$. For instance, in two-party scenario a correlation is called local (Bell used the term \textit{locally-causal}) if it can be expressed as,    
\begin{equation}\label{lhv}
    p(o_1o_2|i_1i_2)=\int_\Lambda p(\lambda) p(o_1|i_1,\lambda) p(o_2|i_2,\lambda) d\lambda,
\end{equation}
where, $\lambda\in\Lambda$ is a random variable, commonly called local hidden variable in quantum foundation community, shared between the two parties following a probability distribution $\{p(\lambda)~|~p(\lambda)\ge0~\&~\int p(\lambda)d\lambda=1\}$ over $\Lambda$. When $\lambda$ takes value from a discrete set, the above integral is replaced by summation. Bell came up with an empirical criterion, famously called the Bell inequality, to test whether a given correlation is local or not. Violation of this inequality establishes nonlocal nature of the correlation, {\it i.e.} the correlation can not be expressed as of Eq.(\ref{lhv}). A correlation is called quantum if it can be obtained through quantum means, {\it i.e.} $P^Q:=\{p(o_1o_2|i_1i_2)\equiv\mbox{Tr}[\rho_{12}(\pi^{i_1}_{o_1}\otimes\pi^{i_2}_{o_2})]\}$, where $\rho_{12}$ is a density operator acting on some composite Hilbert space $\mathcal{H}_1\otimes\mathcal{H}_2$ and $\{\pi^{i_k}_{o_k}\}$ be a positive operator valued measure (POVM) acting on $k^{th}$ parties Hilbert space, {\it i.e.}  $\{\pi^{i_k}_{o_k}\ge0,~\forall~i_k\in\mathcal{I}_k,~o_k\in\mathcal{O}_k,~\&~\sum_{o_k\in\mathcal{O}_k}\pi^{i_k}_{o_k}=\mathbf{I}_k~\forall~i_k\}$, with $\mathbf{I}_k$ being the identity operator on $\mathcal{H}_k$. Let us denote the set of local correlation and quantum correlation as $\mathcal{P}^L$ and $\mathcal{P}^Q$ respectively. While $\mathcal{P}^{NS}$ and $\mathcal{P}^L$ are convex polytope embedded in some Euclidean space, $\mathcal{P}^Q$ is a non-polytopic convex set with infinite, indeed unaccountably many infinite, number of extreme points. We have the following set inclusion relations
\begin{equation}\label{l2}
\mathcal{P}^L_2\subsetneq\mathcal{P}^Q_2\subsetneq\mathcal{P}^{NS}_2. 
\end{equation}
Whereas the proper set inclusion relation $\mathcal{P}^L_2\subsetneq\mathcal{P}^Q_2$ is established through Bell inequality violation, the inclusion relation $\mathcal{P}^Q_2\subsetneq\mathcal{P}^{NS}_2$ is assured from the example of correlation provided by Popescu \& Rohrlich \cite{Popescu_1994}. Whenever cardinality of all the sets $\mathcal{I}_1,\mathcal{I}_2,\mathcal{O}_1$, and $\mathcal{O}_2$ is $2$, Fine \cite{Fine1982} proved that a correlation $P$ will be in $\mathcal{P}^L_2$ {\it if and only if} it satisfies the Clauser-Horne-Shimony-Holt (CHSH) inequality \cite{CHSH:1969}. On the other hand the membership problem to decide whether a given correlation is quantum or not is in general undecidable \cite{slofstra_2019,Slofstra_2019b,ji2020mipre}.  

While in bipartite scenario the set of no-signaling correlations are characterized within local and nonlocal dichotomy, a finer characterization is required when the party number increases \cite{Svet}. 
An $n$-partite input-output correlation is said to be fully local (FL) when
\begin{multline}\label{fl}
  p(o_1\cdots o_N|i_1\cdots i_N)=\int_\Lambda p(\lambda) \Pi_k p(o_k|i_k,\lambda)d\lambda.
\end{multline}
Outside the set $\mathcal{P}^{FL}$ there may exist correlations where nonlocality persists among $m(<n)$ parties but the remaining $n-m$ are locally correlated leading to different new classes of correlations. For instance, in tripartite scenario a correlation is called \textit{bilocal} (BL) if it can be expressed as
\begin{eqnarray}\label{bl}
p(o_1o_2o_3|i_1i_2i_3)=\int p(\lambda) p(o_1|i_1,\lambda)p(o_2o_3|i_2i_3,\lambda)d\lambda.~~~~
\end{eqnarray}
Correlations lying outside the set $\mathcal{P}^{BL}$ are called genuinely nonlocal. Note that, Svetlichny in his paper consider all party permutation while defining a BL correlation. However, here, likewise the Ref.\cite{Gallego2012}, we will restrict ourselves in a particular party permutation, which will make no hindrance in the main purpose of this paper. In multipartite scenario, therefore, the set inclusion relations (\ref{l2}) get modified as, 
\begin{equation}\label{l3}
\mathcal{P}^{FL}\subsetneq\mathcal{P}^{BL},\mathcal{P}^{Q}\subsetneq\mathcal{P}^{NS}. 
\end{equation}
Note that $\mathcal{P}^{BL}$ and $\mathcal{P}^{Q}$ do not follow any subset inclusion relation rather they overlap with each other, {\it i.e.} $\mathcal{P}^{BL}\cap\mathcal{P}^{Q}\neq\mathcal{P}^{BL},\mathcal{P}^{Q},\emptyset$. 

Apart from foundational interest, the characterization of nonlocal correlations is also important from practical prospective as they have shown to be useful resource for device independent quantum information processing \cite{scarani2012}. With this aim several groups have explored the notion of multipartite nonlocality in the recent past \cite{Gallego2012,Barrett2012}. The framework in \cite{Gallego2012} is motivated from the resource theory of quantum entanglement, where entanglement is considered as a useful resource under the operational paradigm of local operation and classical communication (LOCC) \cite{Horodecki-rmp}. In nonlocality scenario, the authors identified the free operations as WCCPI protocols under which the type of nonlocality should not be changed. However, they have pointed out that the a correlation that is local in $1$ vs $23$ ($1|23$) cut can exhibit nonlocality in the same cut after a bona fide WCCPI operation. Such an	inconsistency challenges the completeness of the framework proposed by Svetlichny. The inconsistency stems from the fact that a term like $p(o_2o_3|i_2i_3,\lambda)$ in the decomposition (\ref{bl}) does not need to satisfy the no-signaling constraint. Depending on whether such terms are no-signaling, one-way signaling, or two-way signaling, different sets of correlations can be defined. 
\begin{definition}\label{def1}
(\href{https://doi.org/10.1103/PhysRevLett.109.070401}{PRL {\bf 109}, 070401}) A tripartite correlation $p(o_1o_2o_3|i_1i_2i_3)\equiv P$ is said to admit a time-ordered bilocal (TOBL) model (with respect to the partition $1|23$) if it can be decomposed as,
\begin{subequations}
\label{tobl}
\begin{align}
P=\int p(\lambda)d\lambda p(o_1|i_1,\lambda)p_{2\rightarrow3}(o_2o_3|i_2i_3,\lambda) \\
=\int p(\lambda)d\lambda p(o_1|i_1,\lambda)p_{2\leftarrow3}(o_2o_3|i_2i_3,\lambda), 
\end{align}
\end{subequations}
with the distributions $p_{2\rightarrow3}$ and $p_{2\leftarrow3}$ obeying the conditions
\begin{subequations}
\label{ns}
\begin{align}
p_{2\rightarrow3}(o_2|i_2,\lambda) =\sum_{o_3}p_{2\rightarrow3}(o_2o_3|i_2i_3,\lambda), \\
p_{2\leftarrow3}(o_3|i_3,\lambda) =\sum_{o_2}p_{2\leftarrow3}(o_2o_3|i_2i_3,\lambda). 
\end{align}
\end{subequations}
\end{definition}
The above conditions tell that the term $p(o_2o_3|i_2i_3,\lambda)$ allows only one-way signaling either $2^{nd}$ party to the $3^{rd}$ or vice verse. A tripartite correlation is called NSBL (BL) if the terms $p(o_2o_3|i_2i_3,\lambda)$ obey the NS conditions in both ways (allows two way signalling). The authors in \cite{Gallego2012} have shown that TOBL correlations are consistent with WCCPI protocol along the partition $1|23$, {\it i.e.} such an operation maps TOBL correlations (\ref{tobl}) into a probability distributions with a local model along this partition. They have also reported the following set inclusion relations in the correlation space,
\begin{equation}\label{l4}
\mathcal{P}^{FL}\subsetneq\mathcal{P}^{NSBL}\subsetneq\mathcal{P}^{TOBL}\subsetneq\mathcal{P}^{BL}\subsetneq\mathcal{P}^{NS}. 
\end{equation}
The proper set inclusion relation $\mathcal{P}^{TOBL}\subsetneq\mathcal{P}^{BL}$ has been established by providing an explicit example of quantum correlation in $\mathcal{P}^{BL}$ that exhibits unwanted nonlocal behavior along a particular partition ($1|23$ partition) even under bona fide WCCPI operation. Furthermore it has been claimed that the inconsistency is arising due to the presence of two-way signaling terms in its bi-local decomposition. In fact the authors in \cite{Gallego2012} have made an	observation ``{\it Indeed, all the examples of distributions of the form (4) [also Eq.(\ref{bl}) in our paper] leading to a Bell violation under WCCPI have to be such that the bilocal decomposition requires terms displaying signalling in both directions}". In the following section we will show that this particular sentence needs further clarification as it could be misleading, although the set inclusion relation  (\ref{l4}) is	flawless. However, we will show that a finer classification is possible than the subset inclusion relations of Eq.(\ref{l4}).  

\section{Results}\label{result}
In this section, we first recall the example of correlation provided by Gallego {\it et al.}
They have considered a tripartite quantum correlation obtained from the three-qubit GHZ state: $\ket{\psi_{GHZ}}_{123}=1/\sqrt{2}\left( \ket{000}+\ket{111}\right)_{A_1A_2A_3}$ shared among three parties (say) $A_1,A_2$ and $A_3$. In each run, $A_1$ and $A_2$ perform one of two dichotomous measurements $\mathcal{M}:=\{\sigma_z,\sigma_x\}$, whereas $A_3$ chooses her measurement from  $\mathcal{M}^\prime:=\{(\sigma_z+\sigma_x)/\sqrt{2},(\sigma_z-\sigma_x)/\sqrt{2}\}$. Denoting the first measurement as input $0$ \& the second one as $1$ and $+1$ outcome as $0$ \& $-1$ as $1$ the resulting correlation can be expressed as the following matrix form: 
\begin{equation}\label{g}
	P^G=\frac{1}{2}\left(\begin{array}{cccccccc}
	{2 a^{+}} & {2 a^{-}} & {0} & {0} & {0} & {0} & {2 a^{-}} & {2 a^{+}} \\
	{2 a^{+}} & {2 a^{-}} & {0} & {0} & {0} & {0} & {2 a^{-}} & {2 a^{+}} \\
	{a^{+}} & {a^{-}} & {a^{+}} & {a^{-}} & {a^{-}} & {a^{+}} & {a^{-}} & {a^{+}} \\
	{a^{+}} & {a^{-}} & {a^{+}} & {a^{-}} & {a^{-}} & {a^{+}} & {a^{-}} & {a^{+}} \\
	{a^{+}} & {a^{-}} & {a^{-}} & {a^{+}} & {a^{+}} & {a^{-}} & {a^{-}} & {a^{+}} \\
	{a^{+}} & {a^{-}} & {a^{-}} & {a^{+}} & {a^{+}} & {a^{-}} & {a^{-}} & {a^{+}} \\
	{a^{+}} & {a^{-}} & {a^{-}} & {a^{+}} & {a^{-}} & {a^{+}} & {a^{+}} & {a^{-}} \\
	{a^{-}} & {a^{+}} & {a^{+}} & {a^{-}} & {a^{+}} & {a^{-}} & {a^{-}} & {a^{+}}
	\end{array}\right),
\end{equation}
where $a^{\pm}:=1/4\left(1 \pm 1/\sqrt{2}\right)$. We arrange the input in rows and output in columns and dictionary ordering is followed. This particular correlation is BL as it allows a decomposition (\ref{bl}) across $1|23$ partition and consequently $P^G$ should contain no nonlocal feature in that bi-partition. However, it turns out that after a bona fide WCCPI operation along $1|23$ cut, the resulting correlation $P=\{p(o_1o_3|i_1i_2)\equiv\sum_{o_2} p(o_1o_2o_3|i_1i_2,i_3=o_2)\}$ exhibits CHSH inequality violation. In the required WCCPI protocol, $A_2$ and $A_3$ collaborate in the same laboratory while $A_1$ is in a spatially separated site. After the announcement of the inputs of the nonlocality task, $A_2$ produces her output $o_2$ using the input $i_2$ and then sends it to $A_3$ to use it as input $i_3$, \textit{i.e.}, $i_3=o_2$. Finally, $A_3$ yields output $o_3$. On the other side, $A_1$ locally produces output $o_1$ using input $i_1$. This clearly establishes the incompleteness	of Svetlichny's definitions of bilocality/ genuineness in relation to the operational paradigm of WCCPI. In the next, we show that the correlation $P^G$ allows a bilocal decomposition that contains terms with one-way signaling only. 

\begin{proposition}
The correlation $P^G$ allows a decomposition $P^G=\sum_\lambda p_\lambda p(o_1|i_1,\lambda)p_{2\leftarrow3}(o_2o_3|i_2i_3,\lambda)$, where $p_{2\leftarrow3}(o_2o_3|i_2i_3,\lambda)$ does not admit signaling from $2$ to $3$ but (may) allow signaling from $3$ to $2$.    
\end{proposition}
\begin{widetext}
\begin{proof}
The proof directly follows from the explicit decomposition given by,
\scriptsize
\begin{eqnarray}\label{dg}
P^G=\frac{1}{2}\begin{pmatrix}
	{1} & {0} \\
	{1} & {0}
	\end{pmatrix}_{1} \otimes \begin{pmatrix}
	{a^{+}} & {a^{-}} & {0} & {0} \\
	{a^{+}} & {a^{-}} & {0} & {0} \\
	{a^{+}} & {a^{-}} & {0} & {0} \\
	{0} & {0} & {a^{+}} & {a^{-}}
	\end{pmatrix}_{2 \leftarrow 3}+
	\frac{1}{2}\begin{pmatrix}
	{1} & {0} \\
	{0} & {1}
	\end{pmatrix}_{1} \otimes \begin{pmatrix}
	{a^{+}} & {a^{-}} & {0} & {0} \\
	{a^{+}} & {a^{-}} & {0} & {0} \\
	{0} & {0} & {a^{+}} & {a^{-}} \\
	{a^{+}} & {a^{-}} & {0} & {0}
	\end{pmatrix}_{2 \leftarrow 3} 
	+ \frac{1}{2}\begin{pmatrix}
	{0} & {1} \\
	{1} & {0}
	\end{pmatrix}_{1} \otimes \begin{pmatrix}
	{0} & {0} & {a^{-}} & {a^{+}}\\
	{0} & {0} & {a^{-}} & {a^{+}}\\
	{0} & {0} & {a^{-}} & {a^{+}} \\
	{a^{-}} & {a^{+}} & {0} & {0}
	\end{pmatrix}_{2 \leftarrow 3}+
	\frac{1}{2}\begin{pmatrix}
	{0} & {1} \\
	{0} & {1}
	\end{pmatrix}_{1} \otimes \begin{pmatrix}
	{0} & {0} & {a^{-}} & {a^{+}}\\
	{0} & {0} & {a^{-}} & {a^{+}}\\
	{a^{-}} & {a^{+}} & {0} & {0} \\
	{0} & {0} & {a^{-}} & {a^{+}}
	\end{pmatrix}_{2 \leftarrow 3}.    
\end{eqnarray}
\normalsize
Please note that, the bipartite terms in the decomposition do not allow signaling from $2$ to $3$. 
\end{proof}
\end{widetext}
This proves that a correlation does not require terms displaying signalling in both directions in its bilocal decomposition to show the inconsistent behaviour under WCCPI as claimed in \cite{Gallego2012}. At this point, it is noteworthy that in the above decomposition we have terms that display signaling from $3$ to $2$ whereas the WCCPI protocol used to obtain a bipartite correlation in $1|23$ cut contains signaling from $2$ to $3$. This opposite directional signaling results in the `unwanted' inconsistency. This observation motivates to define an asymmetric version of TOBL correlations.   
\begin{definition}\label{def2}
A tripartite correlation $p(o_1o_2o_3|i_1i_2i_3)$ is said to admit a asymmetric time-ordered bilocal  model from $3$ to $2$ if it allows a decomposition of the form $p(o_1o_2o_3|i_1i_2i_3)=\sum p_\lambda p(o_1|i_1,\lambda)p_{2\leftarrow3}(o_2o_3|i_2i_3,\lambda)$ but need not to allow a decomposition of the form $p(o_1o_2o_3|i_1i_2i_3)=\sum p_\lambda p(o_1|i_1,\lambda)p_{2\rightarrow3}(o_2o_3|i_2i_3,\lambda)$.
\end{definition}
Collection of all such correlations we will denote as $\mathcal{P}^{ATOBL}_{2\leftarrow3}$. Similarly, we can define the set $\mathcal{P}^{ATOBL}_{2\rightarrow3}$. From Definition \ref{def1} \& \ref{def2} arguably it follows that the set 
$\mathcal{P}^{TOBL}$ is intersection of these two asymmetric sets, $\mathcal{P}^{ATOBL}_{2\leftarrow3}\cap\mathcal{P}^{ATOBL}_{2\rightarrow3}=\mathcal{P}^{TOBL}$. Furthermore $\mathcal{P}^{TOBL}$ is a strict subset of both of them. To argue that, first note that the correlation (\ref{g}) [from now on we will denote it as $P^G_{2\leftarrow3}$] does not belong to the set $\mathcal{P}^{ATOBL}_{2\leftarrow3}$, otherwise it will not show the nonlocality across $1|23$ partition under the WCCPI protocol displaying signaling from $3$ to $2$. Since $P^G_{2\leftarrow3}\in\mathcal{P}^{ATOBL}_{2\leftarrow3}$ but $P^G_{2\leftarrow3}\notin\mathcal{P}^{ATOBL}_{2\rightarrow3}$, therefor $P^G_{2\leftarrow3}\notin\mathcal{P}^{TOBL}$ and hence $\mathcal{P}^{TOBL}\subsetneq\mathcal{P}^{ATOBL}_{2\leftarrow3}$. One can obtain a correlation $P^G_{2\rightarrow3}\in\mathcal{P}^{ATOBL}_{2\rightarrow3}$ by just interchanging the measurements for $A_2$ and $A_3$ in $P^G_{2\leftarrow3}$, {\it i.e.} $A_2$ chooses her measurement from the set $\mathcal{M}^\prime$ while $A_3$ from $\mathcal{M}$. To show the nonlocal behaviour of $P^G_{2\rightarrow3}$ in the $1|23$ partition, one have to again interchange the role of $A_2$ and $A_3$ in the WCCPI that has been used for $P^G_{2\leftarrow3}$. Applying similar argument as before it turns out that $\mathcal{P}^{TOBL}\subsetneq\mathcal{P}^{ATOBL}_{2\rightarrow3}$.

We can define a set $\mathcal{P}^{ATOBL}$ of correlations which is the convex hull of $\mathcal{P}^{ATOBL}_{2\leftarrow3}$ and $\mathcal{P}^{ATOBL}_{2\rightarrow3}$, {\it i.e.}, 
\begin{eqnarray}
\mathcal{P}^{ATOBL}&:=\{P~|~P=qP^\prime+(1-q)P^{\prime\prime}\},~~~~~~\\
\mbox{with}~P^\prime,&P^{\prime\prime}\in\mathcal{P}^{ATOBL}_{2\leftarrow3}~\mbox{or}~\mathcal{P}^{ATOBL}_{2\rightarrow3}~\&~q\in[0,1].\nonumber
\end{eqnarray}
One can ask the question whether the set $\mathcal{P}^{ATOBL}$ is same as the set $\mathcal{P}^{ATOBL}_{2\leftarrow3}\cup\mathcal{P}^{ATOBL}_{2\rightarrow3}$. The following proposition answers this question in negative.  
\begin{proposition}\label{prop2}
$\mathcal{P}^{ATOBL}_{2\leftarrow3}\cup\mathcal{P}^{ATOBL}_{2\rightarrow3}\subsetneq\mathcal{P}^{ATOBL}$.
\end{proposition}
\begin{widetext}
\begin{proof}
Consider the following two correlations 
\begin{equation}
\label{kc}
P^\epsilon_{2\leftarrow3}=\frac{1}{2}\left(\begin{array}{cccccccc}	
	{2 k^{+}} & {2 k^{-}} & {0} & {0} & {0} & {0} & {2 k^{-}} & {2 k^{+}} \\
	{2 k^{+}} & {2 k^{-}} & {0} & {0} & {0} & {0} & {2 k^{-}} & {2 k^{+}} \\
	{k^{+}} & {k^{-}} & {k^{+}} & {k^{-}} & {k^{-}} & {k^{+}} & {k^{-}} & {k^{+}} \\
	{k^{+}} & {k^{-}} & {k^{+}} & {k^{-}} & {k^{-}} & {k^{+}} & {k^{-}} & {k^{+}} \\
	{k^{+}} & {k^{-}} & {k^{-}} & {k^{+}} & {k^{+}} & {k^{-}} & {k^{-}} & {k^{+}} \\
	{k^{+}} & {k^{-}} & {k^{-}} & {k^{+}} & {k^{+}} & {k^{-}} & {k^{-}} & {k^{+}} \\
	{k^{+}} & {k^{-}} & {k^{-}} & {k^{+}} & {k^{-}} & {k^{+}} & {k^{+}} & {k^{-}} \\
	{k^{-}} & {k^{+}} & {k^{+}} & {k^{-}} & {k^{+}} & {k^{-}} & {k^{-}} & {k^{+}}
	\end{array}\right),~~~~
P^\epsilon_{2\rightarrow3}=\frac{1}{2}\left(\begin{array}{cccccccc}
	{2 k^{+}} & {0} & {2 k^{-}}  & {0} & {0} & {2 k^{-}} & {0} & {2 k^{+}} \\
	{k^{+}} & {k^{+}} & {k^{-}} & {k^{-}} & {k^{-}} & {k^{-}} & {k^{+}} & {k^{+}} \\
	{2 k^{+}} & {0} & {2 k^{-}}  & {0} & {0} & {2 k^{-}} & {0} & {2 k^{+}} \\
	{k^{+}} & {k^{+}} & {k^{-}} & {k^{-}} & {k^{-}} & {k^{-}} & {k^{+}} & {k^{+}} \\
	{k^{+}} & {k^{-}} & {k^{-}} & {k^{+}} & {k^{+}} & {k^{-}} & {k^{-}} & {k^{+}} \\
	{k^{+}} & {k^{-}} & {k^{-}} & {k^{+}} & {k^{-}} & {k^{+}} & {k^{+}} & {k^{-}} \\	
	{k^{+}} & {k^{-}} & {k^{-}} & {k^{+}} & {k^{+}} & {k^{-}} & {k^{-}} & {k^{+}} \\
	{k^{-}} & {k^{+}} & {k^{+}} & {k^{-}} & {k^{+}} & {k^{-}} & {k^{-}} & {k^{+}}
    \end{array}\right),~~~~
\end{equation}
where, $ k^{\pm}=1/4\left(1 \pm \epsilon \right) $ with $0\leq\epsilon\leq1$. Decomposition, analogous to Eq.(\ref{dg}), for these correlations are given by,
\scriptsize
\begin{subequations}
\begin{align}
P^\epsilon_{2\leftarrow3}&=\frac{1}{2}\begin{pmatrix}
	{1} & {0} \\
	{1} & {0}
	\end{pmatrix} \otimes \begin{pmatrix}
	{k^{+}} & {k^{-}} & {0} & {0} \\
	{k^{+}} & {k^{-}} & {0} & {0} \\
	{k^{+}} & {k^{-}} & {0} & {0} \\
	{0} & {0} & {k^{+}} & {k^{-}}
	\end{pmatrix}+
	\frac{1}{2}\begin{pmatrix}
	{1} & {0} \\
	{0} & {1}
	\end{pmatrix} \otimes \begin{pmatrix}
	{k^{+}} & {k^{-}} & {0} & {0} \\
	{k^{+}} & {k^{-}} & {0} & {0} \\
	{0} & {0} & {k^{+}} & {k^{-}} \\
	{k^{+}} & {k^{-}} & {0} & {0}
	\end{pmatrix} 
	+ \frac{1}{2}\begin{pmatrix}
	{0} & {1} \\
	{1} & {0}
	\end{pmatrix} \otimes \begin{pmatrix}
	{0} & {0} & {k^{-}} & {k^{+}}\\
	{0} & {0} & {k^{-}} & {k^{+}}\\
	{0} & {0} & {k^{-}} & {k^{+}} \\
	{k^{-}} & {k^{+}} & {0} & {0}
	\end{pmatrix}+
	\frac{1}{2}\begin{pmatrix}
	{0} & {1} \\
	{0} & {1}
	\end{pmatrix} \otimes \begin{pmatrix}
	{0} & {0} & {k^{-}} & {k^{+}}\\
	{0} & {0} & {k^{-}} & {k^{+}}\\
	{k^{-}} & {k^{+}} & {0} & {0} \\
	{0} & {0} & {k^{-}} & {k^{+}}
	\end{pmatrix},\\\nonumber\\
P^\epsilon_{2\rightarrow3}&=\frac{1}{2}\begin{pmatrix}
	{1} & {0} \\
	{1} & {0}
	\end{pmatrix} \otimes \begin{pmatrix}
	{k^{+}} & {0} & {k^{-}} & {0} \\
	{k^{+}} & {0} & {k^{-}} & {0} \\
	{k^{+}} & {0} & {k^{-}} & {0} \\
	{0} & {k^{+}} & {0} & {k^{-}}
	\end{pmatrix}+
	\frac{1}{2}\begin{pmatrix}
	{1} & {0} \\
	{0} & {1}
	\end{pmatrix} \otimes \begin{pmatrix}
	{k^{+}} & {0} & {k^{-}} & {0} \\
	{0} & {k^{+}} & {0} & {k^{-}} \\
	{k^{+}} & {0} & {k^{-}} & {0} \\
	{k^{+}} & {0} & {k^{-}} & {0} 
	\end{pmatrix} 
	+ \frac{1}{2}\begin{pmatrix}
	{0} & {1} \\
	{1} & {0}
	\end{pmatrix} \otimes \begin{pmatrix}
	{0} & {k^{-}} & {0} & {k^{+}}\\
	{0} & {k^{-}} & {0} & {k^{+}}\\
	{0} & {k^{-}} & {0} & {k^{+}}\\
	{k^{-}} & {0} & {k^{+}} & {0}
	\end{pmatrix}+
	\frac{1}{2}\begin{pmatrix}
	{0} & {1} \\
	{0} & {1}
	\end{pmatrix} \otimes \begin{pmatrix}
	{0} & {k^{-}} & {0} & {k^{+}}\\
	{k^{-}} & {0} & {k^{+}} & {0} \\
	{0} & {k^{-}} & {0} & {k^{+}}\\	
	{0} & {k^{-}} & {0} & {k^{+}}
	\end{pmatrix}.
\end{align}
\end{subequations}
\normalsize
Consider now another correlation $P^\epsilon_\alpha$ obtained by convex mixing of the correlations in Eq.(\ref{kc}),
\begin{equation}
P^\epsilon_\alpha=\alpha P^\epsilon_{2\leftarrow3}+(1-\alpha)P^\epsilon_{2\rightarrow3}~;~~~\alpha\in[0,1].\end{equation}
As in the case of $P^G$ if we apply the same WCCPI operations containing signaling $2\twoheadleftarrow 3$ and $2\twoheadrightarrow 3$ \cite{note} we will obtain the respective bipartite correlations across the $1|23$ partition: 
\small
\begin{subequations}
\begin{align}
Q_{2\twoheadleftarrow3}(\epsilon,\alpha)&:=\{p(o_1o_3|i_1i_2)\}\equiv\frac{1}{2}
\begin{pmatrix}
	{2k^{+}+\alpha k^-} & {(2-\alpha) k^-}    & {\alpha k^{+}+2(1-\alpha) k^{-}}  & {(2-\alpha) k^{+}+2\alpha k^-} \\
	{2k^{+}+\alpha k^-} & {(2-\alpha) k^-}    & {\alpha k^{+}+2(1-\alpha) k^{-}}  & {(2-\alpha) k^{+}+2\alpha k^-} \\
	{\frac{1}{2}} & {\frac{1}{2}} & {(\alpha +1)k^{+}+(1-\alpha) k^-} & {(1-\alpha) k^{+}+(\alpha +1)k^-} \\
	{2k^{+}}      & {2k^{-}}      & {\alpha k^{+}+(2-\alpha) k^-} & {(2-\alpha) k^{+}+\alpha k^-}
\end{pmatrix},\\\nonumber\\
Q_{2\twoheadrightarrow3}(\epsilon,\alpha)&:=\{p(o_1o_2|i_1i_3)\}\equiv\frac{1}{2}
\begin{pmatrix}
	{2k^{+}+(1-\alpha) k^-} & {(\alpha +1)k^-}    & {(1-\alpha) k^{+}+2\alpha k^{-}}  & {(\alpha +1)k^{+}+2(1-\alpha) k^-} \\
	{2k^{+}+(1-\alpha) k^-} & {(\alpha +1)k^-}    & {(1-\alpha) k^{+}+2\alpha k^{-}}  & {(\alpha +1)k^{+}+2(1-\alpha) k^-} \\
	{\frac{1}{2}} & {\frac{1}{2}} & {(2-\alpha) k^{+}+\alpha k^-} & {\alpha k^{+}+(2-\alpha) k^-} \\
	{2k^{+}}      & {2k^{-}}      & {(1-\alpha) k^{+}+(\alpha +1)k^-} & {(\alpha +1)k^{+}+(1-\alpha) k^-}
\end{pmatrix}.
\end{align}
\end{subequations}
\normalsize
The Bell-CHSH values for these correlations turn out to be 
\begin{align}
\mathcal{B}_{2\twoheadleftarrow3}(\epsilon,\alpha)&=\alpha + \epsilon(3-2\alpha),~~~~~~
\mathcal{B}_{2\twoheadrightarrow 3}(\epsilon,\alpha)=(1-\alpha) + \epsilon(1+2\alpha).
\end{align}
When $P^\epsilon_{2\leftarrow3}$ and $P^\epsilon_{2\rightarrow3}$ are mixed equally, the resulting bipartite correlations obtained under two different WCCPI protocols are same, {\it i.e.} $Q_{2\twoheadleftarrow 3}(\epsilon,1/2)=Q_{2\twoheadrightarrow 3}(\epsilon,1/2):=Q(\epsilon,1/2)$, and consequently their Bell-CHSH values, $\mathcal{B}_{2\twoheadleftarrow 3}(\epsilon,1/2)=\mathcal{B}_{2\twoheadrightarrow 3}(\epsilon,1/2):=\mathcal{B}(\epsilon,1/2)=1/2+2\epsilon$. This leads us to the following conclusion. If the correlation $P^\epsilon_{1/2}$ can be shown to be lie outside the set $\mathcal{P}^{ATOBL}_{2\rightarrow3}$ then it must also lie outside $\mathcal{P}^{ATOBL}_{2\leftarrow3}$. Since $Q(\epsilon,1/2)$ exhibits nonlocality for $\epsilon>3/4$, therefore the correlations $P^{\epsilon>3/4}_{\alpha=1/2}$ (:=$P^{>3/4}_{1/2}$) lie outside the set $\mathcal{P}^{ATOBL}_{2\rightarrow3}\cup\mathcal{P}^{ATOBL}_{2\leftarrow3}$. On the other hand, by construction we have $P^{>3/4}_{1/2}\in\mathcal{P}^{ATOBL}$, which thus implies Proposition \ref{prop2}.
\end{proof}
\begin{figure}[h!]
	\centering
	\includegraphics[width=.70\textwidth]{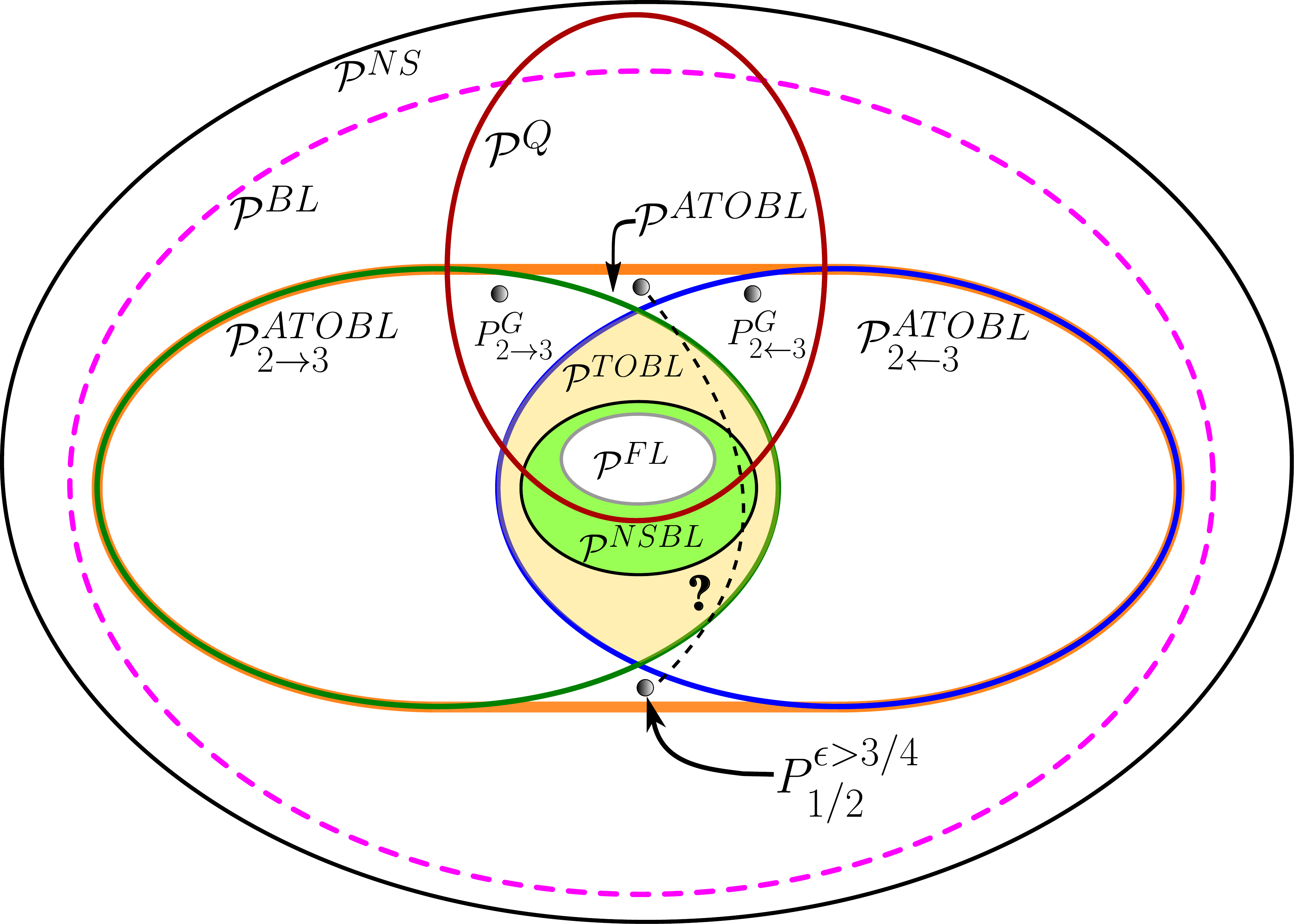}
	\caption{[Color on-line] The inner most white region represents the set of fully local correlations. This and green region represent the set of NSBL correlations. Intersection of the two asymmetric sets $\mathcal{P}^{ATOBL}_{2\leftarrow3}$ and $\mathcal{P}^{ATOBL}_{2\rightarrow3}$ (blue and green elliptical regions, respectively) is the the set of TOBL correlations and their convex hull $\mathcal{P}^{ATOBL}$ is strictly larger than their union. The set $\mathcal{P}^{BL}$ of BL correlations are shown by purple dotted region. It is shown by dotted line as we have not proved $\mathcal{P}^{ATOBL}\subsetneq\mathcal{P}^{BL}$, although we believe the proper inclusion should hold. Red elliptical region $\mathcal{P}^Q$ denotes the set of quantum correlations which strictly contains $\mathcal{P}^{FL}$ but only overlaps with all other sets. For instance the tripartite correlation $P_1\otimes P^{PR}_{23}$ is a non quantum NSBL correlation, where $P$ is some single party input-output probability vector and $P^{PR}$ is the Popescu-Rohrlich correlation. Outermost black ellipse denotes the set of all NS correlations. We put the `?' mark as we are not sure whether the correlation $P^{>3/4}_{1/2}$ is quantum or not. }\label{fig1}
\end{figure}
\end{widetext}
Thus compared to Eq.(\ref{l4}) we have the following finer set inclusion relations in the correlations space.
\begin{eqnarray}
\mathcal{P}^{FL}\subsetneq\mathcal{P}^{NSBL}\subsetneq\mathcal{P}^{TOBL}\subsetneq\mathcal{P}^{ATOBL}_{2\rightarrow3},\mathcal{P}^{ATOBL}_{2\leftarrow3}\nonumber\\
\subsetneq\mathcal{P}^{ATOBL}_{2\rightarrow3}\cup\mathcal{P}^{ATOBL}_{2\leftarrow3}\subsetneq\mathcal{P}^{ATOBL}\subseteq\mathcal{P}^{BL}\subsetneq\mathcal{P}^{NS}.   
\end{eqnarray}
We have not been able to prove the proper subset relation $\mathcal{P}^{ATOBL}\subsetneq\mathcal{P}^{BL}$, although we believe it should hold. Also note that, we have only prove that the correlations $P^{>3/4}_{1/2}$ live in $\mathcal{P}^{ATOBL}$ but not in $\mathcal{P}^{ATOBL}_{2\leftarrow3}\cup\mathcal{P}^{ATOBL}_{2\rightarrow3}$. However, we don't know whether these correlations are quantum or not. The essence of our study can be presented in a Venn diagram as depicted in Fig.\ref{fig1}.

\section{Discussions}\label{discussion}
Bell's theorem addresses one of the long standing debate regarding the foundational status of quantum theory \cite{Einstein:1935,Bohr,Schrodinger:1935,Neumann,Bohm1,Bohm2,gleason1957,jauch1963} and shakes one of our most inveterate world view. Advent of quantum information theory identifies Bell nonlocality as a useful resource for device independent quantum information processing where an information task can be achieved without making any assumptions about the internal working of the devices used in the protocol. Several such protocols have been proposed with many of them already achieving practical implementation \cite{acin2006,Pironio_2010,Colbeck_2012,Brunner,our,our1,chaturvedi,Yin2017}. It has also been established as useful resource in Bayesian game theory to \cite{Brunner:2013,Roy:2016,Banik_2019}. Resource quantification and characterization of such nonlocal correlations is, therefore, relevant from practical point of view. In this paper, we revisit one such framework for multipartite nonlocality developed in Ref.\cite{Gallego2012}. We show that a finer characterization of multipartite nonlocal correlations than that of Ref.\cite{Gallego2012} is possible under the same operational framework proposed there. While doing so we also find some potentially confusing statements made in \cite{Gallego2012} and restate those. Our work accompanying with Ref.\cite{Gallego2012} thus provide a comprehensible picture of multipartite nonlocal correlations.

\begin{acknowledgments}
We thankfully recall many delightful discussions and debates with our colleagues, collaborators, and friends Samir Kunkri, S. Aravinda, Some Sankar Bhattacharya, Arup Roy, Tamal Guha, Som Kanjilal, Debarshi Das, Bihalan Bhattacharya, Ananda G. Maity, Sristy Agrawal, and Sumit Rout. We also gratefully acknowledge private communication with Antonio Ac\'{\i}n. SD acknowledges financial support from INSPIRE-SHE scholarship. MB acknowledges research grant of INSPIRE-faculty fellowship from the Department of Science and Technology, Government of India. 
\end{acknowledgments}


\begin{thebibliography}{35}%
\makeatletter
\providecommand \@ifxundefined [1]{%
 \@ifx{#1\undefined}
}%
\providecommand \@ifnum [1]{%
 \ifnum #1\expandafter \@firstoftwo
 \else \expandafter \@secondoftwo
 \fi
}%
\providecommand \@ifx [1]{%
 \ifx #1\expandafter \@firstoftwo
 \else \expandafter \@secondoftwo
 \fi
}%
\providecommand \natexlab [1]{#1}%
\providecommand \enquote  [1]{``#1''}%
\providecommand \bibnamefont  [1]{#1}%
\providecommand \bibfnamefont [1]{#1}%
\providecommand \citenamefont [1]{#1}%
\providecommand \href@noop [0]{\@secondoftwo}%
\providecommand \href [0]{\begingroup \@sanitize@url \@href}%
\providecommand \@href[1]{\@@startlink{#1}\@@href}%
\providecommand \@@href[1]{\endgroup#1\@@endlink}%
\providecommand \@sanitize@url [0]{\catcode `\\12\catcode `\$12\catcode
  `\&12\catcode `\#12\catcode `\^12\catcode `\_12\catcode `\%12\relax}%
\providecommand \@@startlink[1]{}%
\providecommand \@@endlink[0]{}%
\providecommand \url  [0]{\begingroup\@sanitize@url \@url }%
\providecommand \@url [1]{\endgroup\@href {#1}{\urlprefix }}%
\providecommand \urlprefix  [0]{URL }%
\providecommand \Eprint [0]{\href }%
\providecommand \doibase [0]{http://dx.doi.org/}%
\providecommand \selectlanguage [0]{\@gobble}%
\providecommand \bibinfo  [0]{\@secondoftwo}%
\providecommand \bibfield  [0]{\@secondoftwo}%
\providecommand \translation [1]{[#1]}%
\providecommand \BibitemOpen [0]{}%
\providecommand \bibitemStop [0]{}%
\providecommand \bibitemNoStop [0]{.\EOS\space}%
\providecommand \EOS [0]{\spacefactor3000\relax}%
\providecommand \BibitemShut  [1]{\csname bibitem#1\endcsname}%
\let\auto@bib@innerbib\@empty
\bibitem [{\citenamefont {Bell}(1964)}]{Bell:1964}%
  \BibitemOpen
  \bibfield  {author} {\bibinfo {author} {\bibfnamefont {J.~S.}\ \bibnamefont
  {Bell}},\ }\bibfield  {title} {\enquote {\bibinfo {title} {{On the Einstein
  Podolsky Rosen Paradox}},}\ }\href
  {https://journals.aps.org/ppf/abstract/10.1103/PhysicsPhysiqueFizika.1.195}
  {\bibfield  {journal} {\bibinfo  {journal} {Physics}\ }\textbf {\bibinfo
  {volume} {1}},\ \bibinfo {pages} {195} (\bibinfo {year} {1964})}\BibitemShut
  {NoStop}%
\bibitem [{\citenamefont {Bell}(1966)}]{Bell:1966}%
  \BibitemOpen
  \bibfield  {author} {\bibinfo {author} {\bibfnamefont {J.~S.}\ \bibnamefont
  {Bell}},\ }\bibfield  {title} {\enquote {\bibinfo {title} {{On the Problem of
  Hidden Variables in Quantum Mechanics}},}\ }\href
  {https://journals.aps.org/rmp/abstract/10.1103/RevModPhys.38.447} {\bibfield
  {journal} {\bibinfo  {journal} {Rev. Mod. Phys.}\ }\textbf {\bibinfo {volume}
  {38}},\ \bibinfo {pages} {447} (\bibinfo {year} {1966})}\BibitemShut
  {NoStop}%
\bibitem [{\citenamefont {Brunner}\ \emph {et~al.}(2014)\citenamefont
  {Brunner}, \citenamefont {Cavalcanti}, \citenamefont {Pironio}, \citenamefont
  {Scarani},\ and\ \citenamefont {Wehner}}]{Brunner:2014}%
  \BibitemOpen
  \bibfield  {author} {\bibinfo {author} {\bibfnamefont {N.}~\bibnamefont
  {Brunner}}, \bibinfo {author} {\bibfnamefont {D.}~\bibnamefont {Cavalcanti}},
  \bibinfo {author} {\bibfnamefont {S.}~\bibnamefont {Pironio}}, \bibinfo
  {author} {\bibfnamefont {V.}~\bibnamefont {Scarani}}, \ and\ \bibinfo
  {author} {\bibfnamefont {S.}~\bibnamefont {Wehner}},\ }\bibfield  {title}
  {\enquote {\bibinfo {title} {Bell nonlocality},}\ }\href {\doibase
  10.1103/RevModPhys.86.419} {\bibfield  {journal} {\bibinfo  {journal} {Rev.
  Mod. Phys.}\ }\textbf {\bibinfo {volume} {86}},\ \bibinfo {pages} {419--478}
  (\bibinfo {year} {2014})}\BibitemShut {NoStop}%
\bibitem [{\citenamefont {Mermin}(1993)}]{Mermin1993}%
  \BibitemOpen
  \bibfield  {author} {\bibinfo {author} {\bibfnamefont {N.~D.}\ \bibnamefont
  {Mermin}},\ }\bibfield  {title} {\enquote {\bibinfo {title} {Hidden variables
  and the two theorems of john bell},}\ }\href {\doibase
  10.1103/RevModPhys.65.803} {\bibfield  {journal} {\bibinfo  {journal} {Rev.
  Mod. Phys.}\ }\textbf {\bibinfo {volume} {65}},\ \bibinfo {pages} {803--815}
  (\bibinfo {year} {1993})}\BibitemShut {NoStop}%
\bibitem [{\citenamefont {Svetlichny}(1987)}]{Svet}%
  \BibitemOpen
  \bibfield  {author} {\bibinfo {author} {\bibfnamefont {G.}~\bibnamefont
  {Svetlichny}},\ }\bibfield  {title} {\enquote {\bibinfo {title}
  {Distinguishing three-body from two-body nonseparability by a bell-type
  inequality},}\ }\href {\doibase 10.1103/PhysRevD.35.3066} {\bibfield
  {journal} {\bibinfo  {journal} {Phys. Rev. D}\ }\textbf {\bibinfo {volume}
  {35}},\ \bibinfo {pages} {3066--3069} (\bibinfo {year} {1987})}\BibitemShut
  {NoStop}%
\bibitem [{\citenamefont {Gallego}\ \emph {et~al.}(2012)\citenamefont
  {Gallego}, \citenamefont {W\"urflinger}, \citenamefont {Ac\'{\i}n},\ and\
  \citenamefont {Navascu\'es}}]{Gallego2012}%
  \BibitemOpen
  \bibfield  {author} {\bibinfo {author} {\bibfnamefont {R.}~\bibnamefont
  {Gallego}}, \bibinfo {author} {\bibfnamefont {L.~E.}\ \bibnamefont
  {W\"urflinger}}, \bibinfo {author} {\bibfnamefont {A.}~\bibnamefont
  {Ac\'{\i}n}}, \ and\ \bibinfo {author} {\bibfnamefont {M.}~\bibnamefont
  {Navascu\'es}},\ }\bibfield  {title} {\enquote {\bibinfo {title} {Operational
  framework for nonlocality},}\ }\href {\doibase
  10.1103/PhysRevLett.109.070401} {\bibfield  {journal} {\bibinfo  {journal}
  {Phys. Rev. Lett.}\ }\textbf {\bibinfo {volume} {109}},\ \bibinfo {pages}
  {070401} (\bibinfo {year} {2012})}\BibitemShut {NoStop}%
\bibitem [{\citenamefont {Bancal}\ \emph {et~al.}(2013)\citenamefont {Bancal},
  \citenamefont {Barrett}, \citenamefont {Gisin},\ and\ \citenamefont
  {Pironio}}]{Barrett2012}%
  \BibitemOpen
  \bibfield  {author} {\bibinfo {author} {\bibfnamefont {J.-D.}\ \bibnamefont
  {Bancal}}, \bibinfo {author} {\bibfnamefont {J.}~\bibnamefont {Barrett}},
  \bibinfo {author} {\bibfnamefont {N.}~\bibnamefont {Gisin}}, \ and\ \bibinfo
  {author} {\bibfnamefont {S.}~\bibnamefont {Pironio}},\ }\bibfield  {title}
  {\enquote {\bibinfo {title} {Definitions of multipartite nonlocality},}\
  }\href {\doibase 10.1103/PhysRevA.88.014102} {\bibfield  {journal} {\bibinfo
  {journal} {Phys. Rev. A}\ }\textbf {\bibinfo {volume} {88}},\ \bibinfo
  {pages} {014102} (\bibinfo {year} {2013})}\BibitemShut {NoStop}%
\bibitem [{\citenamefont {Popescu}\ and\ \citenamefont
  {Rohrlich}(1994)}]{Popescu_1994}%
  \BibitemOpen
  \bibfield  {author} {\bibinfo {author} {\bibfnamefont {S.}~\bibnamefont
  {Popescu}}\ and\ \bibinfo {author} {\bibfnamefont {D.}~\bibnamefont
  {Rohrlich}},\ }\bibfield  {title} {\enquote {\bibinfo {title} {Quantum
  nonlocality as an axiom},}\ }\href {\doibase 10.1007/bf02058098} {\bibfield
  {journal} {\bibinfo  {journal} {Found. Phys.}\ }\textbf {\bibinfo {volume}
  {24}},\ \bibinfo {pages} {379--385} (\bibinfo {year} {1994})}\BibitemShut
  {NoStop}%
\bibitem [{\citenamefont {Fine}(1982)}]{Fine1982}%
  \BibitemOpen
  \bibfield  {author} {\bibinfo {author} {\bibfnamefont {A.}~\bibnamefont
  {Fine}},\ }\bibfield  {title} {\enquote {\bibinfo {title} {{Hidden Variables,
  Joint Probability, and the Bell Inequalities}},}\ }\href {\doibase
  10.1103/PhysRevLett.48.291} {\bibfield  {journal} {\bibinfo  {journal} {Phys.
  Rev. Lett.}\ }\textbf {\bibinfo {volume} {48}},\ \bibinfo {pages} {291--295}
  (\bibinfo {year} {1982})}\BibitemShut {NoStop}%
\bibitem [{\citenamefont {Clauser}\ \emph {et~al.}(1969)\citenamefont
  {Clauser}, \citenamefont {Horne}, \citenamefont {Shimony},\ and\
  \citenamefont {Holt}}]{CHSH:1969}%
  \BibitemOpen
  \bibfield  {author} {\bibinfo {author} {\bibfnamefont {J.~F.}\ \bibnamefont
  {Clauser}}, \bibinfo {author} {\bibfnamefont {M.~A.}\ \bibnamefont {Horne}},
  \bibinfo {author} {\bibfnamefont {A.}~\bibnamefont {Shimony}}, \ and\
  \bibinfo {author} {\bibfnamefont {R.~A.}\ \bibnamefont {Holt}},\ }\bibfield
  {title} {\enquote {\bibinfo {title} {{Proposed Experiment to Test Local
  Hidden-Variable Theories}},}\ }\href {\doibase 10.1103/PhysRevLett.23.880}
  {\bibfield  {journal} {\bibinfo  {journal} {Phys. Rev. Lett.}\ }\textbf
  {\bibinfo {volume} {23}},\ \bibinfo {pages} {880--884} (\bibinfo {year}
  {1969})}\BibitemShut {NoStop}%
\bibitem [{\citenamefont {Slofstra}(2019{\natexlab{a}})}]{slofstra_2019}%
  \BibitemOpen
  \bibfield  {author} {\bibinfo {author} {\bibfnamefont {W.}~\bibnamefont
  {Slofstra}},\ }\bibfield  {title} {\enquote {\bibinfo {title} {The set of
  quantum correlations is not closed},}\ }\href {\doibase 10.1017/fmp.2018.3}
  {\bibfield  {journal} {\bibinfo  {journal} {Forum Math. Pi}\ }\textbf
  {\bibinfo {volume} {7}},\ \bibinfo {pages} {e1} (\bibinfo {year}
  {2019}{\natexlab{a}})},\ \bibinfo {note} {(arXiv:1703.08618)}\BibitemShut
  {NoStop}%
\bibitem [{\citenamefont {Slofstra}(2019{\natexlab{b}})}]{Slofstra_2019b}%
  \BibitemOpen
  \bibfield  {author} {\bibinfo {author} {\bibfnamefont {W.}~\bibnamefont
  {Slofstra}},\ }\bibfield  {title} {\enquote {\bibinfo {title} {Tsirelson's
  problem and an embedding theorem for groups arising from non-local games},}\
  }\href {\doibase 10.1090/jams/929} {\bibfield  {journal} {\bibinfo  {journal}
  {J. Am. Math. Soc}\ }\textbf {\bibinfo {volume} {33}},\ \bibinfo {pages}
  {1--56} (\bibinfo {year} {2019}{\natexlab{b}})},\ \bibinfo {note}
  {(arXiv:1606.03140)}\BibitemShut {NoStop}%
\bibitem [{\citenamefont {Ji}\ \emph {et~al.}(2020)\citenamefont {Ji},
  \citenamefont {Natarajan}, \citenamefont {Vidick}, \citenamefont {Wright},\
  and\ \citenamefont {Yuen}}]{ji2020mipre}%
  \BibitemOpen
  \bibfield  {author} {\bibinfo {author} {\bibfnamefont {Z.}~\bibnamefont
  {Ji}}, \bibinfo {author} {\bibfnamefont {A.}~\bibnamefont {Natarajan}},
  \bibinfo {author} {\bibfnamefont {T.}~\bibnamefont {Vidick}}, \bibinfo
  {author} {\bibfnamefont {J.}~\bibnamefont {Wright}}, \ and\ \bibinfo {author}
  {\bibfnamefont {H.}~\bibnamefont {Yuen}},\ }\href@noop {} {\enquote {\bibinfo
  {title} {{MIP*=RE}},}\ } (\bibinfo {year} {2020}),\ \Eprint
  {http://arxiv.org/abs/2001.04383} {arXiv:2001.04383 [quant-ph]} \BibitemShut
  {NoStop}%
\bibitem [{\citenamefont {Scarani}(2012)}]{scarani2012}%
  \BibitemOpen
  \bibfield  {author} {\bibinfo {author} {\bibfnamefont {V.}~\bibnamefont
  {Scarani}},\ }\bibfield  {title} {\enquote {\bibinfo {title} {The
  device-independent outlook on quantum physics},}\ }\href
  {http://www.physics.sk/aps/pubs/2012/aps-12-04/aps-12-04.pdf} {\bibfield
  {journal} {\bibinfo  {journal} {Acta Phys. Slovaca}\ }\textbf {\bibinfo
  {volume} {62}},\ \bibinfo {pages} {347--409} (\bibinfo {year}
  {2012})}\BibitemShut {NoStop}%
\bibitem [{\citenamefont {Horodecki}\ \emph {et~al.}(2009)\citenamefont
  {Horodecki}, \citenamefont {Horodecki}, \citenamefont {Horodecki},\ and\
  \citenamefont {Horodecki}}]{Horodecki-rmp}%
  \BibitemOpen
  \bibfield  {author} {\bibinfo {author} {\bibfnamefont {R.}~\bibnamefont
  {Horodecki}}, \bibinfo {author} {\bibfnamefont {P.}~\bibnamefont
  {Horodecki}}, \bibinfo {author} {\bibfnamefont {M.}~\bibnamefont
  {Horodecki}}, \ and\ \bibinfo {author} {\bibfnamefont {K.}~\bibnamefont
  {Horodecki}},\ }\bibfield  {title} {\enquote {\bibinfo {title} {{Quantum
  entanglement}},}\ }\href {\doibase 10.1103/RevModPhys.81.865} {\bibfield
  {journal} {\bibinfo  {journal} {Rev. Mod. Phys.}\ }\textbf {\bibinfo {volume}
  {81}},\ \bibinfo {pages} {865--942} (\bibinfo {year} {2009})}\BibitemShut
  {NoStop}%
\bibitem [{not()}]{note}%
  \BibitemOpen
  \href@noop {} {\ }\bibinfo {note} {Here, double arrow `$2\twoheadleftarrow
  3$' denotes WCCPI operation with communication from $A_3$ to $A_2$ as
  described previously, while `$2\twoheadrightarrow 3$' implies the
  opposite.}\BibitemShut {Stop}%
\bibitem [{\citenamefont {Einstein}\ \emph {et~al.}(1935)\citenamefont
  {Einstein}, \citenamefont {Podolsky},\ and\ \citenamefont
  {Rosen}}]{Einstein:1935}%
  \BibitemOpen
  \bibfield  {author} {\bibinfo {author} {\bibfnamefont {A.}~\bibnamefont
  {Einstein}}, \bibinfo {author} {\bibfnamefont {B.}~\bibnamefont {Podolsky}},
  \ and\ \bibinfo {author} {\bibfnamefont {N.}~\bibnamefont {Rosen}},\
  }\bibfield  {title} {\enquote {\bibinfo {title} {{Can Quantum-Mechanical
  Description of Physical Reality Be Considered Complete?}}}\ }\href {\doibase
  10.1103/PhysRev.47.777} {\bibfield  {journal} {\bibinfo  {journal} {Phys.
  Rev.}\ }\textbf {\bibinfo {volume} {47}},\ \bibinfo {pages} {777--780}
  (\bibinfo {year} {1935})}\BibitemShut {NoStop}%
\bibitem [{\citenamefont {Bohr}(1935)}]{Bohr}%
  \BibitemOpen
  \bibfield  {author} {\bibinfo {author} {\bibfnamefont {N.}~\bibnamefont
  {Bohr}},\ }\bibfield  {title} {\enquote {\bibinfo {title} {Can
  quantum-mechanical description of physical reality be considered complete?}}\
  }\href {\doibase 10.1103/PhysRev.48.696} {\bibfield  {journal} {\bibinfo
  {journal} {Phys. Rev.}\ }\textbf {\bibinfo {volume} {48}},\ \bibinfo {pages}
  {696--702} (\bibinfo {year} {1935})}\BibitemShut {NoStop}%
\bibitem [{\citenamefont {Schr\"{o}dinger}(1935)}]{Schrodinger:1935}%
  \BibitemOpen
  \bibfield  {author} {\bibinfo {author} {\bibfnamefont {E.}~\bibnamefont
  {Schr\"{o}dinger}},\ }\bibfield  {title} {\enquote {\bibinfo {title}
  {{Discussion of Probability Relations between Separated Systems}},}\ }\href
  {https://doi.org/10.1017/S0305004100013554} {\bibfield  {journal} {\bibinfo
  {journal} {Proc. Cambridge Philos. Soc.}\ }\textbf {\bibinfo {volume} {31}},\
  \bibinfo {pages} {553} (\bibinfo {year} {1935})}\BibitemShut {NoStop}%
\bibitem [{Neu()}]{Neumann}%
  \BibitemOpen
  \href@noop {} {\ }\bibinfo {note} {{J. von Neumann, ``Mathematishe Grundlagen
  der Quanten-mechanik", Verlag Julius-Springer, Berlin (1932), [English
  translation: Princeton University Press (1955)]}}\BibitemShut {NoStop}%
\bibitem [{\citenamefont {Bohm}(1952{\natexlab{a}})}]{Bohm1}%
  \BibitemOpen
  \bibfield  {author} {\bibinfo {author} {\bibfnamefont {D.}~\bibnamefont
  {Bohm}},\ }\bibfield  {title} {\enquote {\bibinfo {title} {{A Suggested
  Interpretation of the Quantum Theory in Terms of "Hidden" Variables. I}},}\
  }\href {\doibase 10.1103/PhysRev.85.166} {\bibfield  {journal} {\bibinfo
  {journal} {Phys. Rev.}\ }\textbf {\bibinfo {volume} {85}},\ \bibinfo {pages}
  {166--179} (\bibinfo {year} {1952}{\natexlab{a}})}\BibitemShut {NoStop}%
\bibitem [{\citenamefont {Bohm}(1952{\natexlab{b}})}]{Bohm2}%
  \BibitemOpen
  \bibfield  {author} {\bibinfo {author} {\bibfnamefont {D.}~\bibnamefont
  {Bohm}},\ }\bibfield  {title} {\enquote {\bibinfo {title} {{A Suggested
  Interpretation of the Quantum Theory in Terms of "Hidden" Variables. II}},}\
  }\href {\doibase 10.1103/PhysRev.85.180} {\bibfield  {journal} {\bibinfo
  {journal} {Phys. Rev.}\ }\textbf {\bibinfo {volume} {85}},\ \bibinfo {pages}
  {180--193} (\bibinfo {year} {1952}{\natexlab{b}})}\BibitemShut {NoStop}%
\bibitem [{\citenamefont {Gleason}(1957)}]{gleason1957}%
  \BibitemOpen
  \bibfield  {author} {\bibinfo {author} {\bibfnamefont {A.~M.}\ \bibnamefont
  {Gleason}},\ }\bibfield  {title} {\enquote {\bibinfo {title} {{Measures on
  the closed subspaces of a Hilbert space}},}\ }\href@noop {} {\bibfield
  {journal} {\bibinfo  {journal} {J. Math. Mech.}\ }\textbf {\bibinfo {volume}
  {6}},\ \bibinfo {pages} {885--893} (\bibinfo {year} {1957})}\BibitemShut
  {NoStop}%
\bibitem [{\citenamefont {Jauch}\ and\ \citenamefont
  {Piron}(1963)}]{jauch1963}%
  \BibitemOpen
  \bibfield  {author} {\bibinfo {author} {\bibfnamefont {J.~M.}\ \bibnamefont
  {Jauch}}\ and\ \bibinfo {author} {\bibfnamefont {C.}~\bibnamefont {Piron}},\
  }\bibfield  {title} {\enquote {\bibinfo {title} {Can hidden variables be
  excluded in quantum mechanics},}\ }\href
  {https://cds.cern.ch/record/345316/files/CM-P00056869.pdf} {\bibfield
  {journal} {\bibinfo  {journal} {Helv. Phys. Acta}\ }\textbf {\bibinfo
  {volume} {36}},\ \bibinfo {pages} {827--837} (\bibinfo {year}
  {1963})}\BibitemShut {NoStop}%
\bibitem [{\citenamefont {Ac\'{\i}n}\ \emph {et~al.}(2006)\citenamefont
  {Ac\'{\i}n}, \citenamefont {Gisin},\ and\ \citenamefont
  {Masanes}}]{acin2006}%
  \BibitemOpen
  \bibfield  {author} {\bibinfo {author} {\bibfnamefont {A.}~\bibnamefont
  {Ac\'{\i}n}}, \bibinfo {author} {\bibfnamefont {N.}~\bibnamefont {Gisin}}, \
  and\ \bibinfo {author} {\bibfnamefont {L.}~\bibnamefont {Masanes}},\
  }\bibfield  {title} {\enquote {\bibinfo {title} {{From Bell's Theorem to
  Secure Quantum Key Distribution}},}\ }\href {\doibase
  10.1103/PhysRevLett.97.120405} {\bibfield  {journal} {\bibinfo  {journal}
  {Phys. Rev. Lett.}\ }\textbf {\bibinfo {volume} {97}},\ \bibinfo {pages}
  {120405} (\bibinfo {year} {2006})}\BibitemShut {NoStop}%
\bibitem [{\citenamefont {Pironio}\ \emph {et~al.}(2010)\citenamefont
  {Pironio}, \citenamefont {Ac{\'{\i}}n}, \citenamefont {Massar}, \citenamefont
  {de~la Giroday}, \citenamefont {Matsukevich}, \citenamefont {Maunz},
  \citenamefont {Olmschenk}, \citenamefont {Hayes}, \citenamefont {Luo},
  \citenamefont {Manning},\ and\ \citenamefont {Monroe}}]{Pironio_2010}%
  \BibitemOpen
  \bibfield  {author} {\bibinfo {author} {\bibfnamefont {S.}~\bibnamefont
  {Pironio}}, \bibinfo {author} {\bibfnamefont {A.}~\bibnamefont
  {Ac{\'{\i}}n}}, \bibinfo {author} {\bibfnamefont {S.}~\bibnamefont {Massar}},
  \bibinfo {author} {\bibfnamefont {A.~Boyer}\ \bibnamefont {de~la Giroday}},
  \bibinfo {author} {\bibfnamefont {D.~N.}\ \bibnamefont {Matsukevich}},
  \bibinfo {author} {\bibfnamefont {P.}~\bibnamefont {Maunz}}, \bibinfo
  {author} {\bibfnamefont {S.}~\bibnamefont {Olmschenk}}, \bibinfo {author}
  {\bibfnamefont {D.}~\bibnamefont {Hayes}}, \bibinfo {author} {\bibfnamefont
  {L.}~\bibnamefont {Luo}}, \bibinfo {author} {\bibfnamefont {T.~A.}\
  \bibnamefont {Manning}}, \ and\ \bibinfo {author} {\bibfnamefont
  {C.}~\bibnamefont {Monroe}},\ }\bibfield  {title} {\enquote {\bibinfo {title}
  {{Random numbers certified by Bell's theorem}},}\ }\href {\doibase
  10.1038/nature09008} {\bibfield  {journal} {\bibinfo  {journal} {Nature}\
  }\textbf {\bibinfo {volume} {464}},\ \bibinfo {pages} {1021--1024} (\bibinfo
  {year} {2010})}\BibitemShut {NoStop}%
\bibitem [{\citenamefont {Colbeck}\ and\ \citenamefont
  {Renner}(2012)}]{Colbeck_2012}%
  \BibitemOpen
  \bibfield  {author} {\bibinfo {author} {\bibfnamefont {R.}~\bibnamefont
  {Colbeck}}\ and\ \bibinfo {author} {\bibfnamefont {R.}~\bibnamefont
  {Renner}},\ }\bibfield  {title} {\enquote {\bibinfo {title} {Free randomness
  can be amplified},}\ }\href {\doibase 10.1038/nphys2300} {\bibfield
  {journal} {\bibinfo  {journal} {Nat. Phys.}\ }\textbf {\bibinfo {volume}
  {8}},\ \bibinfo {pages} {450--453} (\bibinfo {year} {2012})}\BibitemShut
  {NoStop}%
\bibitem [{\citenamefont {Brunner}\ \emph {et~al.}(2008)\citenamefont
  {Brunner}, \citenamefont {Pironio}, \citenamefont {Ac\'{\i}n}, \citenamefont
  {Gisin}, \citenamefont {M\'ethot},\ and\ \citenamefont {Scarani}}]{Brunner}%
  \BibitemOpen
  \bibfield  {author} {\bibinfo {author} {\bibfnamefont {N.}~\bibnamefont
  {Brunner}}, \bibinfo {author} {\bibfnamefont {S.}~\bibnamefont {Pironio}},
  \bibinfo {author} {\bibfnamefont {A.}~\bibnamefont {Ac\'{\i}n}}, \bibinfo
  {author} {\bibfnamefont {N.}~\bibnamefont {Gisin}}, \bibinfo {author}
  {\bibfnamefont {A.~A.}\ \bibnamefont {M\'ethot}}, \ and\ \bibinfo {author}
  {\bibfnamefont {V.}~\bibnamefont {Scarani}},\ }\bibfield  {title} {\enquote
  {\bibinfo {title} {{Testing the Dimension of Hilbert Spaces}},}\ }\href
  {\doibase 10.1103/PhysRevLett.100.210503} {\bibfield  {journal} {\bibinfo
  {journal} {Phys. Rev. Lett.}\ }\textbf {\bibinfo {volume} {100}},\ \bibinfo
  {pages} {210503} (\bibinfo {year} {2008})}\BibitemShut {NoStop}%
\bibitem [{\citenamefont {Das}\ \emph {et~al.}(2013)\citenamefont {Das},
  \citenamefont {Banik}, \citenamefont {Rai}, \citenamefont {Gazi},\ and\
  \citenamefont {Kunkri}}]{our}%
  \BibitemOpen
  \bibfield  {author} {\bibinfo {author} {\bibfnamefont {S.}~\bibnamefont
  {Das}}, \bibinfo {author} {\bibfnamefont {M.}~\bibnamefont {Banik}}, \bibinfo
  {author} {\bibfnamefont {A.}~\bibnamefont {Rai}}, \bibinfo {author}
  {\bibfnamefont {Md.~R.}\ \bibnamefont {Gazi}}, \ and\ \bibinfo {author}
  {\bibfnamefont {S.}~\bibnamefont {Kunkri}},\ }\bibfield  {title} {\enquote
  {\bibinfo {title} {Hardy's nonlocality argument as a witness for postquantum
  correlations},}\ }\href {\doibase 10.1103/PhysRevA.87.012112} {\bibfield
  {journal} {\bibinfo  {journal} {Phys. Rev. A}\ }\textbf {\bibinfo {volume}
  {87}},\ \bibinfo {pages} {012112} (\bibinfo {year} {2013})}\BibitemShut
  {NoStop}%
\bibitem [{\citenamefont {Mukherjee}\ \emph {et~al.}(2015)\citenamefont
  {Mukherjee}, \citenamefont {Roy}, \citenamefont {Bhattacharya}, \citenamefont
  {Das}, \citenamefont {Gazi},\ and\ \citenamefont {Banik}}]{our1}%
  \BibitemOpen
  \bibfield  {author} {\bibinfo {author} {\bibfnamefont {A.}~\bibnamefont
  {Mukherjee}}, \bibinfo {author} {\bibfnamefont {A.}~\bibnamefont {Roy}},
  \bibinfo {author} {\bibfnamefont {S.~S.}\ \bibnamefont {Bhattacharya}},
  \bibinfo {author} {\bibfnamefont {S.}~\bibnamefont {Das}}, \bibinfo {author}
  {\bibfnamefont {Md.~R.}\ \bibnamefont {Gazi}}, \ and\ \bibinfo {author}
  {\bibfnamefont {M.}~\bibnamefont {Banik}},\ }\bibfield  {title} {\enquote
  {\bibinfo {title} {Hardy's test as a device-independent dimension witness},}\
  }\href {\doibase 10.1103/PhysRevA.92.022302} {\bibfield  {journal} {\bibinfo
  {journal} {Phys. Rev. A}\ }\textbf {\bibinfo {volume} {92}},\ \bibinfo
  {pages} {022302} (\bibinfo {year} {2015})}\BibitemShut {NoStop}%
\bibitem [{\citenamefont {Chaturvedi}\ and\ \citenamefont
  {Banik}(2015)}]{chaturvedi}%
  \BibitemOpen
  \bibfield  {author} {\bibinfo {author} {\bibfnamefont {A.}~\bibnamefont
  {Chaturvedi}}\ and\ \bibinfo {author} {\bibfnamefont {M.}~\bibnamefont
  {Banik}},\ }\bibfield  {title} {\enquote {\bibinfo {title}
  {Measurement-device--independent randomness from local entangled states},}\
  }\href {https://doi.org/10.1209/0295-5075/112/30003} {\bibfield  {journal}
  {\bibinfo  {journal} {EPL}\ }\textbf {\bibinfo {volume} {112}},\ \bibinfo
  {pages} {30003} (\bibinfo {year} {2015})}\BibitemShut {NoStop}%
\bibitem [{\citenamefont {et~al.}(2017)}]{Yin2017}%
  \BibitemOpen
  \bibfield  {author} {\bibinfo {author} {\bibfnamefont {J.~Yin}\ \bibnamefont
  {et~al.}},\ }\bibfield  {title} {\enquote {\bibinfo {title} {Satellite-based
  entanglement distribution over 1200 kilometers},}\ }\href {\doibase
  10.1126/science.aan3211} {\bibfield  {journal} {\bibinfo  {journal}
  {Science}\ }\textbf {\bibinfo {volume} {356}},\ \bibinfo {pages} {1140--1144}
  (\bibinfo {year} {2017})}\BibitemShut {NoStop}%
\bibitem [{\citenamefont {Brunner}\ and\ \citenamefont
  {Linden}(2013)}]{Brunner:2013}%
  \BibitemOpen
  \bibfield  {author} {\bibinfo {author} {\bibfnamefont {N.}~\bibnamefont
  {Brunner}}\ and\ \bibinfo {author} {\bibfnamefont {N.}~\bibnamefont
  {Linden}},\ }\bibfield  {title} {\enquote {\bibinfo {title} {{Connection
  between Bell nonlocality and Bayesian game theory}},}\ }\href
  {http://www.nature.com/articles/ncomms3057} {\bibfield  {journal} {\bibinfo
  {journal} {Nat. Comm.}\ }\textbf {\bibinfo {volume} {4}},\ \bibinfo {pages}
  {2057} (\bibinfo {year} {2013})}\BibitemShut {NoStop}%
\bibitem [{\citenamefont {Roy}\ \emph {et~al.}(2016)\citenamefont {Roy},
  \citenamefont {Mukherjee}, \citenamefont {Guha}, \citenamefont {Ghosh},
  \citenamefont {Bhattacharya},\ and\ \citenamefont {Banik}}]{Roy:2016}%
  \BibitemOpen
  \bibfield  {author} {\bibinfo {author} {\bibfnamefont {A.}~\bibnamefont
  {Roy}}, \bibinfo {author} {\bibfnamefont {A.}~\bibnamefont {Mukherjee}},
  \bibinfo {author} {\bibfnamefont {T.}~\bibnamefont {Guha}}, \bibinfo {author}
  {\bibfnamefont {S.}~\bibnamefont {Ghosh}}, \bibinfo {author} {\bibfnamefont
  {S.~S.}\ \bibnamefont {Bhattacharya}}, \ and\ \bibinfo {author}
  {\bibfnamefont {M.}~\bibnamefont {Banik}},\ }\bibfield  {title} {\enquote
  {\bibinfo {title} {{Nonlocal correlations: Fair and unfair strategies in
  Bayesian games}},}\ }\href {\doibase 10.1103/PhysRevA.94.032120} {\bibfield
  {journal} {\bibinfo  {journal} {Phys. Rev. A}\ }\textbf {\bibinfo {volume}
  {94}},\ \bibinfo {pages} {032120} (\bibinfo {year} {2016})}\BibitemShut
  {NoStop}%
\bibitem [{\citenamefont {Banik}\ \emph {et~al.}(2019)\citenamefont {Banik},
  \citenamefont {Bhattacharya}, \citenamefont {Ganguly}, \citenamefont {Guha},
  \citenamefont {Mukherjee}, \citenamefont {Rai},\ and\ \citenamefont
  {Roy}}]{Banik_2019}%
  \BibitemOpen
  \bibfield  {author} {\bibinfo {author} {\bibfnamefont {M.}~\bibnamefont
  {Banik}}, \bibinfo {author} {\bibfnamefont {S.~S.}\ \bibnamefont
  {Bhattacharya}}, \bibinfo {author} {\bibfnamefont {N.}~\bibnamefont
  {Ganguly}}, \bibinfo {author} {\bibfnamefont {T.}~\bibnamefont {Guha}},
  \bibinfo {author} {\bibfnamefont {A.}~\bibnamefont {Mukherjee}}, \bibinfo
  {author} {\bibfnamefont {A.}~\bibnamefont {Rai}}, \ and\ \bibinfo {author}
  {\bibfnamefont {A.}~\bibnamefont {Roy}},\ }\bibfield  {title} {\enquote
  {\bibinfo {title} {Two-qubit pure entanglement as optimal social welfare
  resource in bayesian game},}\ }\href {\doibase 10.22331/q-2019-09-09-185}
  {\bibfield  {journal} {\bibinfo  {journal} {Quantum}\ }\textbf {\bibinfo
  {volume} {3}},\ \bibinfo {pages} {185} (\bibinfo {year} {2019})}\BibitemShut
  {NoStop}%
\end{thebibliography}

%

\end{document}